\documentclass[
    ,final            
  ,draft            
  ,numberedheadings 
  ]
  {aipproc}

\usepackage[mathscr]{eucal}
 \usepackage{amsmath, amssymb}

\def\dmf{\dot{\mathfrak{M}}}
 
  \def\sun{\hbox{$\odot$}}
   
\newcommand{\be}{\begin{equation}}
\newcommand{\ee}{\end{equation}}
\newcommand{\bdm}{\begin{displaymath}}
\newcommand{\edm}{\end{displaymath}}
\layoutstyle{6x9}

\begin{document}

\noindent {\it Accepted for publication in Astronomy Reports, vol.\,59, No.5
(2015)}

\vspace{1cm}

\title{Period clustering of the anomalous X-ray pulsars}


\author{G.S.\,Bisnovatyi-Kogan}{
  address={Space Research Institute of RAS,  84/32 Profsoyuznaya Str, Moscow 117997, Russia, and \\
National Research Nuclear University ``MEPhI'', Kashirskoye shosse 31, Moscow 115409,
Russia} }

\author{N.R.\,Ikhsanov}{
  address={Pulkovo Observatory, Pulkovskoe Shosse 65, Saint-Petersburg 196140, Russia, and \\
  Saint Petersburg State University,  Universitetsky pr., 28, Saint Petersburg 198504, Russia}
}

\begin{abstract}
In this paper we address the question of why the observed periods  of the Anomalous
X-ray Pulsars (AXPs) and Soft Gamma-ray Repeaters (SGRs) are clustered in the range
2--12\,s. We explore a possibility to answer this question assuming that AXPs and SGRs
are the descendants of High Mass X-ray Binaries (HMXBs) which have been disintegrated in
the core-collapse supernova explosion. The spin period of neutron stars in HMXBs evolves
towards the equilibrium period, averaging around a few seconds.  After
the explosion of its massive companion, the neutron star turns out to be embedded into a
dense gaseous envelope, the accretion from which leads to the formation of a residual
magnetically levitating (ML) disk. We show that the expected mass of a disk in this case
is $10^{-7} - 10^{-8}\,{\rm M_{\sun}}$ which is sufficient to maintain the process of
accretion at the rate $10^{14} - 10^{15}$\,g/s over a  time span of a few thousand
years. During this period the star manifests itself as an isolated X-ray pulsar with
a number of parameters resembling those of AXPs and SGRs. Period clustering of such pulsars
can be provided if the lifetime of the residual disk does not exceed the spin-down
timescale of the neutron star.
\end{abstract}

\maketitle


   \section{Introduction}

In the previous paper \cite{Bisnovatyi-Kogan-Ikhsanov-2014} we have explored the possibility
of modelling  the spin evolution and X-ray emission of Anomalous X-ray Pulsars (AXPs) and Soft Gamma-ray
Repeaters (SGRs)  within a scenario of magnetic-levitation accretion
\cite{Ikhsanov-etal-2014}. We have considered a magnetized
isolated neutron star accreting material from a residual, slowly rotating
magnetically-levitating disk (ML-disk).
Using the parameters of this process estimated in \cite{Shvartsman-1971,
Bisnovatyi-Kogan-Ruzmaikin-1974, Bisnovatyi-Kogan-Ruzmaikin-1976,
Igumenshchev-etal-2003, Ikhsanov-Beskrovnaya-2012, Ikhsanov-Finger-2012,
Ikhsanov-etal-2013}, we came to a conclusion that the basic features of the  X-ray emission from AXPs and SGRs
 can be explained by the  magnetic-levitation accretion  model
 without an assumption about super-strong dipole magnetic field of
these neutron stars to be invoked. In particular, the magnetic field strength on the surface of AXPs and SGRs,
evaluated within this approach to fit their observed spin-down rates, is in the range
$B_{\rm s} \sim (0.01 - 10) \times 10^{12}$\,G, and the expected black-body temperature
of the pulsar X-ray radiation  $kT \sim 0.3 - 1.4$\,keV,
which is close to the observed value \cite{Bisnovatyi-Kogan-Ikhsanov-2014}.

One of the possible mechanisms of  flaring activity of these objects in
gamma-rays mentioned in our previous paper is a spontaneous release of energy
accumulated in a non-equilibrium
layer of super-heavy nuclei located at the lower boundary of the upper crust of a low-
or moderate-mass neutron star, suggested in  \cite{Bisnovatyi-Kogan-Chechetkin-1974,
Bisnovatyi-Kogan-etal-1975, Bisnovatyi-Kogan-Chechetkin-1979}.
In the frame of this approach,  the AXPs and SGRs can be
distinguished by a low mass of the neutron stars that provides  the maximum amount of nuclear power in their
crust for the minimum energy necessary to trigger its release.
The formation of such objects is likely to be a rather rare event, which
can account for their relatively low fraction (about 1\%) in the family of neutron stars
\cite{Vigano-etal-2013, Olausen-Kaspi-2014}.

At the same time, observational data on the radio-pulsar
J\,1518-4904, composing a close binary system with another neutron star, favor a possibility for such objects to exist.
The pulsar mass measured at the confidence level of  95.4\% is
$m_p=0.72^{+0.51}_{-0.58} M_\odot$, while its companion has a mass of
$m_c=2.00^{+0.58}_{-0.51} M_\odot$ \cite{Tauris-etal-2013}. In the previous paper
 \cite{Bisnovatyi-Kogan-Ikhsanov-2014} we have already mentioned that
 a light neutron star can be formed due to off-center nuclear outburst induced
 by the core-collapse of the star in the process of supernova explosion.
 This scenario was studied in  \cite{Branch-Nomoto-1986},
presenting the modeling of the SN1b explosion in terms of collapse of the stellar remnant
consisting of the iron core of the mass of $0.4 - 0.6\,M_{\sun}$,
surrounded by the helium and carbon-oxygen shells. It has been noted that
a rapid temperature rise during the collapse of such a remnant leads to the thermonuclear explosion
and, finally, to the ejection of both shells.  However, up to now no detailed simulations of this scenario
have been performed.

In this paper we discuss a period clustering of AXPs and SGRs in a relatively
narrow interval from 2 to 12 seconds. Analysis of this phenomenon is
traditionally performed under assumption about a relatively young age
 of these objects ($\sim 10^2 - 10^5$\,years), which is based on the estimates of their
 spin-down timescale, $\tau_{\rm sd} \sim
P_{\rm s}/2 \dot{P}$, and/or of the age of supernova remnants associated with
them. Here $P_{\rm s}$ is the spin period of a pulsar and $\dot{P} = dP_{\rm
s}/dt$ is its current spin-down rate. In this case, the model of rotational evolution
of AXPs and SGRs can be constructed by making additional assumptions that
either these neutron stars have initially super-strong and rapidly decaying dipole magnetic fields
(see \cite{Colpi-etal-2000, Pons-etal-2013} and references therein), or
they prove to be in the state of exceptionally intensive fall-back accretion
just after their birth (see \cite{Ertan-etal-2009} and references therein).

In both scenarios, there is only remote possibility  to find an AXP with the period
significantly smaller than 1\,s, since
the characteristic time over which the  spin period  of a neutron star rises from an initially
very short value up to a few seconds is significantly smaller than $\tau_{\rm sd}$.
A probability to discover a star with a spin period significantly in excess of the
observed values is also negligible provided either the decay time of super-strong magnetic field
(in the magnetar model) or the life-time of a residual disk in the model of fall-back accretion
does not exceed  $\tau_{\rm sd}$. Thus, in the frame of these scenarios
the observed period clustering of AXPs is caused mainly by the selection effect.
In this context, the neutron star acquires the capacity to display gamma-ray flares
exclusively at the moment of birth and gradually loses this ability on the timescale
 $\sim \tau_{\rm sd}$.

In our previous paper \cite{Bisnovatyi-Kogan-Ikhsanov-2014}, we have attempted
to apply the above approach to  study the spin evolution of a neutron star undergoing
fall-back accretion from a non-Keplerian magnetic disk.
We have pointed out that it is difficult to explain the observed period clustering of AXPs
in the frame of this approach without additional assumptions which essentially
narrow the range of possible scenarios. This has stimulated us to search for alternative
solutions, one of which is discussed in this paper.

A suggested scenario  is based on the hypothesis that  AXPs and
SGRs are  descendants of the neutron stars which had been X-ray pulsars in High Mass X-ray Binaries
(HMXBs)during a previous epoch. The reason for this is provided by a simple fact that
information on the characteristic spin-down timescale of the pulsar, $\tau_{\rm sd}$, or on the age of
a supernova remnant in which it is located, is not sufficient for correct evaluation of
its genuine age. This information can only allow to assert that
the time interval from the star transit into the state of regular spin-down
(in which it is observed in the present epoch) does not exceed  $\tau_{\rm sd}$. The true age of the star
can, however, substantially exceed $\tau_{\rm sd}$.

As an example of such situation one can consider the disintegration of a HMXB
due to the core-collapse of its massive
component in the process of supernova explosion.  The old neutron
star which had evolved as a member of a HMXB during the main sequence evolution of  its massive
companion  ($10^6 -10^7$\,years), after disruption of the binary  becomes isolated  and
 embedded into a supernova remnant of significantly younger age.  Under the conditions of interest
 its spin period at the moment of system disintegration is about a few seconds (see Section\,\ref{eqp}).
Capturing matter from the envelope ejected by its exploded companion, this star can find itself
in the state of an isolated X-ray pulsar accreting material onto its surface from a residual disk
(see Section \,\ref{capture}). Main features of such sources are briefly discussed and compared
with observed parameters of AXPs and SGRs  in
Section\,\ref{discussion}.

 \section{Pulsar period in a High-Mass X-ray Binary}\label{eqp}

The majority of presently known X-ray pulsars are the members of HMXBs, which
are close pairs consisting of a massive O/B-star and a neutron star with strong
magnetic field. Their X-ray emission is generated due to accretion of matter
onto the neutron star surface in the magnetic polar regions. The period of a
pulsar corresponds to the spin period of a neutron star,
 $P_{\rm s}$, and its luminosity, $L = \dmf GM_{\rm
ns}/R_{\rm ns}$, is determined by the rate of mass accretion onto the stellar surface,
$\dmf$, where $M_{\rm ns}$ is the mass and  $R_{\rm ns}$ is the radius of a neutron
star. The period of such a pulsar evolves according to the equation
 \be\label{main}
 2 \pi I \dot{\nu} = K_{\rm su} - K_{\rm sd}
 \ee
towards the equilibrium period, $P_{\rm eq}$, whose value is defined by the balance of
spin-up, $K_{\rm su}$, and spin-down, $K_{\rm sd}$, torques, exerted on  the star by the
accretion flow. Here  $I$ is the moment of inertia of a neutron star,  $\nu = 1/P_{\rm
s}$ is the frequency of its axial rotation and  $\dot{\nu} = d\nu/dt$.

According to modern views, HMXBs are the descendants of high-mass binaries,
which survived  a  supernova explosion caused by a core-collapse of their more
massive component. A neutron star born in the course of this event (the first
supernova explosion) possesses strong magnetic field and rotates with a period
of a fraction of a second. Then it is spinning down because of magneto-dipole
losses (the ejector state) and later on due to interaction between its magnetic
field and the surrounding plasma (the propeller state). When its spin period
reaches the critical value determined by the equality of its corotation radius,
$r_{\rm cor} = \left(GM_{\rm ns}/\omega_{\rm s}^2\right)^{1/3}$, and the
magnetosphere radius, $r_{\rm m}$, the star passes to the accretor state and
manifests itself as an X-ray pulsar. Here $\omega_{\rm s} = 2 \pi/P_{\rm s}$ is
the angular velocity of its axial rotation.

Numerical simulations of this scenario \cite{Urpin-etal-1998} indicate that a
neutron star with a sufficiently strong initial magnetic field at the final
 stage of its evolution in a HMXB undergoes accretion from a Keplerian disk and
its spin period evolves towards the equilibrium period whose value can be
estimated as
  \be
 P_{\rm eq}^{\rm (Kd)} \simeq 3\,{\rm s}\ \times\ k_{\rm t}^{1/2}\ \mu_{30}^{6/7}\ m^{5/7}\ \dmf_{17}^{-3/7}.
 \ee
Here $k_{\rm t}$ is a dimensionless parameter of the order of unity, $\mu_{30}$ is the
dipole magnetic moment of the neutron star in units of
$10^{30}\,\text{G\,$\cdot\,$cm$^3$}$, $m$ is the neutron star mass in  units of
$1.4\,M_{\sun}$ and $\dmf_{17}$ is the mass accretion rate onto the stellar surface in
units of  $10^{17}$\,g/s. The age of the neutron star by this moment is
 \be\label{tms}
 t_{\rm ms} \simeq 6 \times 10^6\ \left(\frac{M_{\rm opt}}{20\,{\rm M_{\sun}}}\right)^{-5/2}\ {\rm yr},
 \ee
that corresponds to the average time of its massive component (of the mass $M_{\rm
opt}$) evolution on the main sequence  \cite{Bhattacharya-van-den-Heuvel-1991}.

As the massive star begins to evolve off of the main sequence, its radius increases and it
fills its Roche lobe. In this case the mass exchange between the system components
proceeds on the thermal timescale, $t_{\rm th} = GM_{\rm opt}^2/(R_{\rm opt} L_{\rm
opt}) \sim 3 \times 10^7\,m^2$\,yr, in the form of a stream through the Lagrangian point
L1 at the rate $\dmf \sim M_{\rm opt}/t_{\rm th} \simeq 10^{17}\,(M_{\rm
opt}/20\,M_{\sun})^{-1}$\,g/s. Here $R_{\rm opt}$ and $L_{\rm opt}$ are the radius and
the luminosity of the massive component determined in general case by its mass  (see
\cite{Masevich-Tutukov-1988} and references therein). For the stars with the mass
$M_{\rm opt} \sim 12 - 40\,M_{\sun}$, the rate of mass exchange between the system
components falls in a relatively narrow interval  $\dmf \sim (0.6 - 2) \times
10^{17}$\,g/s. Thus, the main parameter determining the spin period of the neutron star
at the final stage of the HMXB evolution,  $P_{\rm eq}^{\rm (Kd)}$, is the strength of its dipole magnetic field.

\begin{figure}
\includegraphics[scale=0.6]{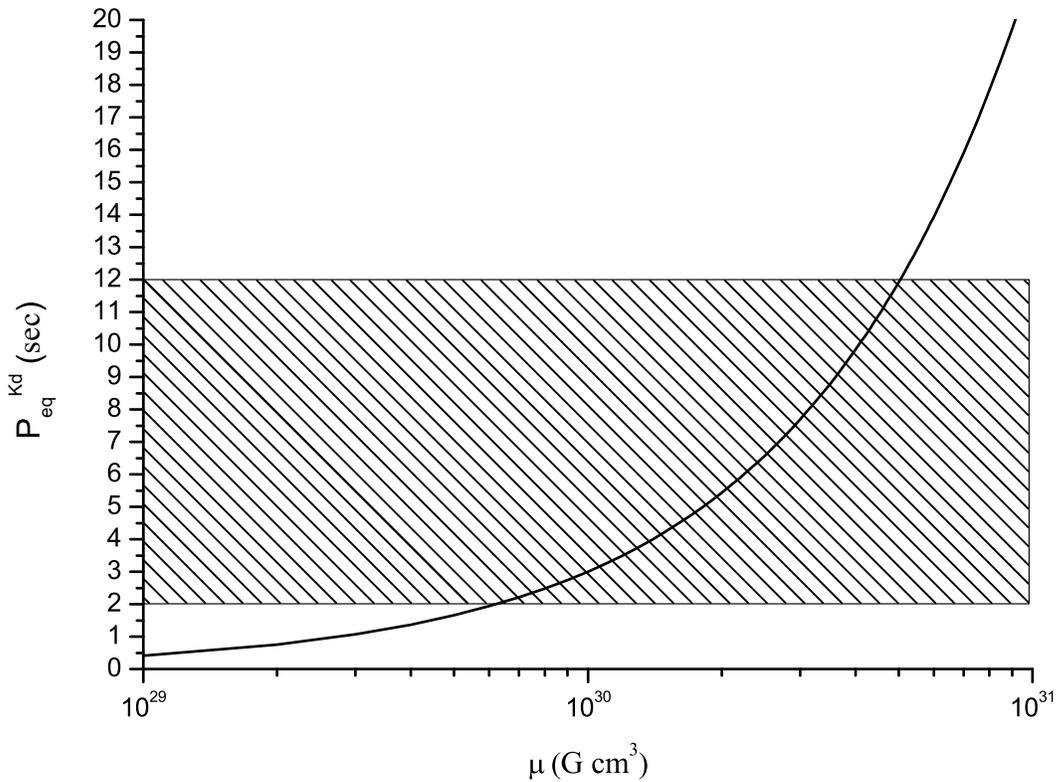}
\caption{Equilibrium period of a light ($0.8\,M_{\sun}$) neutron star accreting material from a
Keplerian disk at the final stage of evolution of a high-mass X-ray binary in
dependence of its dipole magnetic
moment.  Dashed region covers observed periods of AXPs and
SGRs.}
   \label{fig-1}
 \end{figure}

 Dependence $P_{\rm eq}^{\rm (Kd)} = P_{\rm eq}^{\rm (Kd)}(\mu)$ for a light
($0.8\,M_{\sun}$) neutron star accreting material from a Keplerian disk at the average rate
$10^{17}\,{\rm g\,s^{-1}}$ is presented in the Figure. Dashed region shows
 the observed period range of AXPs and SGRs. As can be seen in the figure, the spin period of a neutron star
at the final stage of a HMXB evolution falls into the dashed region provided
its dipole magnetic moment is in the range $(1 - 5) \times 10^{30}\,\text{G\,cm$^3$}$.
For the average stellar radius of 12\,km
 \cite{Potekhin-2010} this corresponds to the dipole  field strength of
 $10^{12} - 10^{13}$\,G on the stellar surface in the region of its magnetic poles.

The absolute majority of presently known neutron stars possess the magnetic field within the above mentioned limits.
This is valid for both the isolated radio-pulsars and X-ray pulsars in the HMXBs, in which the surface magnetic field was estimated through observations of the cyclotron line in their X-ray spectra
(see \cite{Caballero-Wilms-2012, Baushev-Bisnovatyi-Kogan-1999} and references therein).
The periods of these pulsars at the final stage of a HMXB evolution
 may tend to clustering around an average value, $P_{\rm eq}^{\rm (Kd)}$,
   which is of the order of a few seconds for the parameters of interest.

However, it should be mentioned that the magnetic field strength of only 7 out of 15
pulsars considered in the previous paper
\cite{Bisnovatyi-Kogan-Ikhsanov-2014}, fall within above mentioned interval.
According to our estimates, the lower boundary of this interval exceeds the lower limit
to the magnetic field strength of the remaining eight X-ray sources
(\cite{Bisnovatyi-Kogan-Ikhsanov-2014}, Table\,3). This can be connected with oversimplification
of our model  in which the rotation velocity of matter in the residual magnetically-levitating disk
was adopted to be zero. One cannot totally exclude a possibility of enhanced dissipation of the neutron star
magnetic field and/or
change of its configuration during the period of its flaring activity.
This can be, in particular, stimulated by the anomalous heating of the neutron star crust, if the source of its magnetic field is located relatively close to its surface. Finally, it is worthwhile to mention
a large diversity of possible scenarios of the final stage of the HMXBs evolution, and to
notice that realization of our hypothesis is expected predominantly within a compact model of pre-supernova,
which was used for interpretation of the SN\,1987A
\cite{Imshennik-Nadyozhin-1988}.


 \section{Formation of a residual disk}\label{capture}

Evolution of the binary system at the phase of a HMXB ends up with a core-collapse of its
massive component accompanied with a supernova explosion and ejection of a massive shell
($1-3\,M_{\sun}$). As a result, the second neutron star or a black hole is born.
Depending on the energetics and geometry of the explosion, a HMXB can  either become a
system of two degenerate objects or disintegrate into two isolated compact stars with
the  latter case being more probable  \cite{Popov-Prokhorov-2006}. The characteristic
time  of system disintegration,  $\tau_{\rm dec} \sim a/v_{\rm kick}$, is determined by
the initial orbital separation, $a$, and the value of their spatial velocity, $v_{\rm kick}$,
achieved in the process of supernova explosion.

Because of the envelope ejection in the supernova explosion, both stars turn out
to be embedded into a dense gaseous medium. The velocity of the main part of the ejecta
at the initial phase of its expansion significantly exceeds  $1\,000$\,km/s
and reaches  10\,000\,km/s in the outermost layers. This powerful plasma flow interacts
with the old neutron star and disrupts the accretion structure formed around its magnetosphere
in the previous epoch (e.g. a Keplerian accretion disk) before final dispersion  into an
extended supernova remnant.
However, the initial expansion velocity of the inner portion of the ejecta is significantly smaller.
Calculations \cite{Michel-1988, Chevalier-1989} have shown that the innermost parts of the ejecta with
the mass under $10^{-5}\,M_{\sun}$ remain gravitationally bound with the young compact object, and
soon after the supernova explosion return to a new-born star in the form of fall-back accretion flow.
The mass of this portion of matter which is in the state of free expansion with a relatively slow velocity
$\sim 100-1000$\,km/s does not exceed $M_0 \sim
10^{-3}\,M_{\sun}$ \cite{Imshennik-Nadyozhin-1988}.
However, it is this material which can form   a massive
magnetically-levitating disk surrounding the magnetosphere of the old neutron star. Accreting from this disk,
the star shows itself as an isolated X-ray
pulsar with the period $P_{\rm eq}^{\rm (Kd)}$. In this Section we show that the life-time of such a pulsar can be as long
as a few thousand years.

The amount of mass with which an old neutron star interacts in a unit time moving with a
relative velocity   $v_{\rm rel}$ can be estimated as follows
 \be
 \dmf_{\rm cap} = \pi r_{\!_{\rm G}}^2 \rho_{\rm env} v_{\rm rel} \simeq
 10^{20}\,{\rm g\,s^{-1}} \times m^2 v_8^{-3} a_{13}^{-3} \left(\frac{M_0}{10^{-3}\,M_{\sun}}\right).
 \ee
Here $r_{\!_{\rm G}} = 2GM_{\rm ns}/v_{\rm rel}^2$ is the Bondi radius of the
neutron star, $\rho_{\rm env} \sim 3 M_0/\left(4 \pi a^3\right)$ is the mean
density of the material in the inner part of the envelope with the mass $M_0$
contained inside radius $a$, $v_8 = v_{\rm rel}/10^8$\,cm and $a_{13} =
a/10^{13}$\,cm.

The total amount of matter which can under favorable conditions be captured by
the neutron star over the time span  $\tau_{\rm dec}$ from this part of the
envelope is
 \be
 M_{\rm cap} \simeq 10^{-8}\,M_{\sun}\ \times\ m^2 a_{13}^{-2} v_8^{-3}
 \left(\frac{M_0}{10^{-3}\,{\rm M_{\sun}}}\right)
 \left(\frac{v_{\rm kick}}{300\,\text{km/s}}\right)^{-1}.
 \ee
The parameter $v_{\rm kick}$ in this expression is normalized to its average value derived in  \cite{Lyne-Lorimer-1994,
Arzoumanian-etal-2002}.

The structure of an accretion flow forming by the captured material inside Bondi radius
is determined by the relative velocity of the star, $v_{\rm rel}$, as well as by
physical parameters of the gas and magnetic field strength in the surrounding envelope.
As has been recently shown by  Ikhsanov et\,al. (see \cite{Ikhsanov-etal-2014a} and
references therein), a scenario of quasi-spherical accretion can be realized under the
condition  $v_{\rm rel} > v_{\rm ma}$, where
   \be\label{vma}
 v_{\rm ma} \simeq 3000\,{\rm km\,s^{-1}} \times \beta_0^{-1/5}\ \mu_{30}^{-6/35}\
 \dmf_{20}^{3/35}\ m^{12/35}\ c_7^{2/5}.
 \ee
Here $\beta_0 = E_{\rm th}(r_{\!_{\rm G}})/E_{\rm m}(r_{\!_{\rm G}})$ is the ratio of
the thermal, $E_{\rm th} \sim \rho c_{\rm s}^2$, to magnetic, $E_{\rm m} = B_{\rm f}^2/8
\pi$, energy in the material captured by the star at its Bondi radius. Parameters
$\rho$, $B_{\rm f}$ and $c_{\rm s}$ denote the density, the magnetic field strength and
the sound speed in the accretion flow, respectively. If $v_{\rm rel} < v_{\rm kd}$,
where
  \be
v_{\rm kd} \simeq 30\,{\rm km\,s^{-1}} \times
\xi_{0.2}^{3/7}\,\beta_0^{1/7}\,m^{3/7}\,c_7^{-2/7}\,\left(\frac{P_{\rm
orb}}{100\,{\rm days}}\right)^{-3/7},
 \ee
the material captured by the star is accumulated in a Keplerian accretion disk
\cite{Bisnovatyi-Kogan-Ikhsanov-2014}. Here $P_{\rm orb}$ is the orbital period of the
HMXB and $\xi_{0.2} = \xi/0.2$ is a parameter accounting for the angular momentum
dissipation in the quasi-spherical non-magnetic flow, normalized on its average value
obtained in \cite{Ruffert-1999}.

Finally, in the intermediate case  $v_{\rm kd} < v_{\rm rel} < v_{\rm ma}$, we expect
realization of magnetic-levitation accretion scenario in which the matter captured by
the neutron star is accumulated around its magnetosphere in a form of non-Keplerian
magnetically levitating disk  and moves towards the star in the diffusion regime. The
outer radius of the ML-disk is determined by the Shvartsman radius
\cite{Shvartsman-1971},
 \be\label{rsh}
 R_{\rm sh} = \beta_0^{-2/3} r_{\!_{\rm G}} \left(\frac{c_{\rm s}(r_{\!_{\rm G}})}{v_{\rm rel}}\right)^{4/3},
 \ee
 at which the magnetic pressure in the free-falling gas reaches its ram
pressure. In the presence of large-scale magnetic field with the inhomogeneity
scale in excess of Bondi radius, the initially quasi-spherical flow rapidly
decelerates and transforms its geometry into a ML-disk
\cite{Bisnovatyi-Kogan-Ruzmaikin-1974, Bisnovatyi-Kogan-Ruzmaikin-1976}.

The inner radius of a ML-disk corresponds to the magnetosphere radius of the
neutron star and equals   \cite{Ikhsanov-etal-2013, Ikhsanov-etal-2014}
 \be\label{rma}
 r_{\rm ma} = \left(\frac{c\,m_{\rm p}^2}{16\,\sqrt{2}\,e\,k_{\rm B}}\right)^{2/13}
 \frac{\alpha_{\!_{\rm B}}^{2/13} \mu^{6/13} (GM_{\rm ns})^{5/13}}{T_0^{2/13} L_{\rm X}^{4/13} R_{\rm ns}^{4/13}}.
 \ee
Here $m_{\rm p}$ and $e$ are  the proton mass and the electron charge, and $k_{\rm B}$
is the Boltzmann constant. $T_0$ is the gas temperature in the diffusion layer at the
magnetosphere boundary (magnetopause) , and
 $\alpha_{\!_{\rm B}} $ is a dimensionless parameter, expressing the ratio
 of the effective coefficient of the accretion
flow diffusion into the magnetic field of the star, $D_{\rm eff}$, to the Bohm
diffusion coefficient.

The mass of a ML-disk forming in this scenario is
 \be\label{mdisk}
 M_{\rm d} = 4 \pi \int\limits_{r_{\rm ma}}^{R_{\rm sh}} \rho(r) h_{\rm z}(r) r dr,
 \ee
where $\rho(r)$ is the density of the disk, and $h_{\rm z}(r)$ is its
half-thickness.These parameters can be estimated taking into account that the
gaseous (as well as the magnetic) pressure in the ML-disk reaches its maximum,
  \be
 \rho(r_{\rm ma}) c_{\rm s}^2(r_{\rm ma}) = \frac{\mu^2}{2 \pi r_{\rm ma}^6},
 \ee
at the inner radius of the disk and decreases with distance from the star as $\rho(r)
c_{\rm s}^2(r) \propto r^{-5/2}$ \cite{Bisnovatyi-Kogan-Ruzmaikin-1974,
Bisnovatyi-Kogan-Ruzmaikin-1976}. Taking into account that the gas temperature in the
disk is
 \be\label{tdisk}
 T(r) = \left(\frac{\dmf GM_{\rm ns}}{4 \pi r^3 \sigma_{\!_{\rm SB}}}\right)^{1/4},
 \ee
and, correspondingly, the sound speed is $c_{\rm s} \sim \left(k_{\rm B} T/m_{\rm
p}\right)^{1/2} \propto r^{-3/8}$, the density distribution in the radial direction is
 \be
 \rho(r) = \rho(r_{\rm ma}) \left(\frac{r}{r_{\rm ma}}\right)^{-7/4},
 \ee
where $\sigma_{\!_{\rm SB}}$ is the Stefan-Boltzmann constant. Finally, the
half-thickness of the disk can be estimated as
\cite{Bisnovatyi-Kogan-Ruzmaikin-1974, Bisnovatyi-Kogan-Ruzmaikin-1976}
 \be\label{hz}
 h_{\rm z}(r) = \left(\frac{k_{\rm B} T(r) r^3}{m_{\rm p} GM_{\rm ns}}\right)^{1/2}.
 \ee

Substituting  (\ref{rma}), (\ref{rsh}) and (\ref{tdisk}--\ref{hz}) to
(\ref{mdisk}) and taking into account that under the conditions of interest
$R_{\rm sh} \gg r_{\rm ma}$, we find
 \be
 M_{\rm d} \simeq 10^{-7}\,{\rm M_{\sun}} \times \alpha_{0.1}^{-7/3} \beta_0^{-11/12}
 \mu_{30}^{5/13}\ \dmf_{20}^{99/104}\ m^{25/52} c_7^{11/6} v_8^{-55/12},
 \ee
where parameter  $\alpha_{0.1} = \alpha_{\!_{\rm B}}/0.1$ is normalized
following \cite{Gosling-etal-1991}. Thus, if an old neutron star captures material from the
envelope having been ejected in the collapse of its massive companion, this can result
in formation of a ML-disk with the mass sufficient to provide the process of accretion at the rate $10^{14} -
10^{15}$\,g/s over a time span of a few thousand years. Within this scenario, the ML-disk is composed of the matter
located in the inner parts of the ejecta where the expansion velocity does not exceed 1000\,km/s.

 \section{Conclusions}\label{discussion}

We show that isolated X-ray pulsars with the periods of a few seconds can be
descendants of the HMXBs. Within our scenario, these objects are old ($10^6 - 10^7$\,years)
neutron stars with the dipole magnetic field in the range $10^{10} - 10^{13}$\,G, which
accrete matter onto their surface from a residual slowly rotating ML-disk. Under these conditions,
the spin periods of the pulsars are increasing at a high rate and their  life-time  is
limited by  that of a residual disk. If this time does not exceed the spin-down timescale
of a neutron star, their periods are expected to cluster around an average value of a few seconds.
If the life-time of a residual disk is smaller than the typical dissipation time of a supernova remnant,
these neutron stars can be embedded into the nebulosity forming in the explosion of their
massive  HMXB companion, which resulted in the disintegration of the binary system during the previous epoch.

It is quite noticeable that the objects described above are similar to AXPs and SGRs
 in some of their manifestations. In particular, the periods of isolated X-ray pulsars
 which are descendants of HMXBs are tending to clustering around an average value
   which is of the order of a few seconds for the classical range of the neutron stars
   magnetic fields  ($10^{12} -
10^{13}$\,G). At the same time, these stars are regularly spinning down at a rather high rate
due to accretion of matter from the ML-disk onto their surface. The sources arising in this scenario
can be situated in the supernova remnants in spite of the fact that their age exceeds by an order of magnitude the average life-time of such nebulosities. On the other hand, the epoch during which a neutron star
is a member of a HMXB is long enough for its spin period to increase from an arbitrarily small
initial value to a few seconds. Finally, as was first mentioned by  Mereghetti \& Stella
 \cite{Mereghetti-Stella-1995}, the model of accretion onto a neutron star provides the simplest
 explanation of the spectral characteristics of the X-ray emission observed from AXPs and SGRs (see, also,
\cite{van-Paradijs-etal-1995, Marsden-etal-2001, Truemper-etal-2010, Truemper-etal-2013,
Bisnovatyi-Kogan-Ikhsanov-2014}).

An attempt to apply our scenario in order to associate AXPs with the descendants of HMXBs
rises, however,  a number of questions. First, within this approach one cannot eliminate a possibility
that in the vicinity of AXP there exists a young compact supernova remnant formed in the core-collapse
of the massive HMXB component. Discovery of such a source or its contribution to the energetics and/or
structure of the nebulosity could help to estimate the prospects of a hypothesis that AXPs are descendants of HMXBs.
Another question remaining open is why an old neutron star exhibits activity in the gamma-rays only after
disintegration of the HMXB and merely during the phase of accretion from the ML-disk.

In the previous paper \cite{Bisnovatyi-Kogan-Ikhsanov-2014} we have mentioned that
accretion of matter onto the stellar surface could be considered as one of possible triggers
of the flares. On the other hand,  the star was also in the state of accretion during its evolution
as a part of the HMXB. One of the factors discriminating these phases of accretion could be the chemical
composition of matter captured by the star from its surroundings. In particular, after the supernova explosion
the accretion flow can be expected to be enriched with heavy elements.
Accumulation of this material on the stellar surface may cause a thermonuclear flare which exceeds in power
the super-flares in X-ray bursters
 \cite{Cooper-etal-2009,
Potekhin-Chabrier-2012, Stevens-etal-2014} and can trigger reactions leading to the release of nuclear energy
stored  in the non-equilibrium layer of the neutron star crust. The degree of impact which this factor
 would have on the ability of neutron stars to produce gamma-ray flares will be addressed in one of the upcoming papers.

\begin{theacknowledgments}
The authors thank an anonymous referee for constructive criticism and stimulating comments, as well as
N.G.\,Beskrovnaya and V.Yu.\,Kim  for useful discussions and
help in preparation of this manuscript. The work was partly supported by RFBR
under the grants No.\,13-02-00077 and No.\,14-02-00728, SPbSU under the grant
No.\,6.38.18.2014, the President Support Program for Leading Scientific
Schools NSH-261.2014.2, and the RAS Presidium Program No.\,41.
\end{theacknowledgments}

\end{document}